\documentstyle[twocolumn,aps]{revtex}

\input epsf.sty

\begin{document}

\wideabs{
\draft
\title{Geometric relation between centrality and the impact parameter in
relativistic heavy-ion collisions \cite{grant}}
\author{Wojciech Broniowski and Wojciech Florkowski}
\address{The H. Niewodnicza\'nski Institute of Nuclear Physics,
PL-31342 Cracow, Poland}
\maketitle
\begin{abstract}
We show, under general assumptions which
are well satisfied in relativistic heavy-ion
collisions, that the geometric relation of centrality $c$
to the impact parameter $b$, namely
$c \simeq \pi b^2/\sigma_{\rm inel}$, holds to a very high accuracy
for all but most peripheral collisions. More precisely, if $c(N)$
is the centrality of events with the multiplicity higer than $N$,
then $b$ is the value of the impact parameter for which the average
multiplicity $\bar{n}(b)$ is equal to $N$. 
The corrections to this geometric formula are of the order 
$(\Delta n(b)/\bar{n}(b))^2$, where $\Delta n(b)$ is 
the width of the multiplicity distribution at a given value of $b$, hence
are very small. In other words, the
centrality effectively measures the impact parameter.
\end{abstract}
}

Data from relativistic heavy-ion collisions (SPS, RHIC) are typically
categorized by introducing {\em centrality}, $c$, defined as the percentile
of events with the largest number of produced particles (as registered in
detectors), or largest number of participants (as determined from
zero-degree calorimeters). We denote this number generically as $n$. Results
of measurements, such as multiplicities \cite{phoboscentr,brahms}, $p_{\perp
}$ spectra \cite{velko,harris,QM}, the elliptic flow coefficient $v_{2}$ 
\cite{starelfl,starielfl}, the HBT radii \cite{starhbt}, {\em etc.}, are
then presented for various centralities. From the experimental viewpoint the
centrality is a good, unambiguous criterion allowing to divide the data. On
the other hand, theoreticians need to assign an impact parameter, $b$, to a
given centrality. The impact parameter is in a sense more basic, since it
determines the initial geometry of the collision and appears across the
formalism. Theoretical calculations in heavy-ion physics input $b$ in order
to obtain predictions. Having done the calculation, the question arises as
to which centrality data should the model results be compared to. For that
purpose one typically applies the Glauber model in order to compute the
number of wounded nucleons or binary collisions at a given $b$, which are
subsequently related to multiplicities or number of participants \cite
{khhet,huo}. Since these are measured in the experiment, one is able to
identify $b$ with $c$.

In this paper we argue that such an effort is not necessary, since, under
general assumptions which hold very well in relativistic heavy-ion
collisions, we have, to a very high precision, the relation 
\begin{equation}
c(N)\simeq \frac{\pi b(N)^{2}}{\sigma _{{\rm inel}}},\qquad {\rm for\ }b<%
\bar{R}  \label{rel}
\end{equation}
where $\sigma _{{\rm inel}}$ is the total inelastic nucleus-nucleus cross
section, and $\bar{R}$ is of the order of the sum of the radii of the
colliding nuclei. The centrality $c(N)$ is the centrality of events with the
multiplicity higher than $N$, while $b(N)$ is the value of the impact
parameter for which the average multiplicity $\bar{n}(b)$ is equal to $N$.
As will be shown, Eq. (\ref{rel}) holds to a high accuracy for all but
most peripheral collisions. Note that it is geometric in nature, and does
not involve explicitly the variable $n$ needed to categorize the data
(multiplicities, number of participants, number of binary collisions, {\em %
etc.}). At first glance, this fact may seem a bit surprising.

One can explain the geometric nature of (\ref{rel}), and the fact that it
does not explicitly depend on $n$, with the following pedagogical example.
Consider a competition where archers are shooting at a target of radius $R$,
each of them once. The archers are very poor, such that they shoot randomly.
They are paid accordingly to their aim: more central, higher reward. We are
not allowed to watch the competition, hence do not know which spot on the
target has been hit, but later we review the reward records. Suppose a large
number of archers scored (here we take only 10 in order to write down the
results explicitly), and are ranked according to their prizes, which are:
100\$, 100\$, 50\$, 50\$, 50\$, 10\$, 10\$, 10\$, 10\$, 10\$. The two
archers that got the highest prize (100\$ in this case) had to hit the
bull's eye. Since these are the 20\% of all archers, and they were shooting
randomly, we can immediately determine (neglecting the statistical error)
the radius $b$ of the bull's eye, since 20\% is the ratio of the area of the
bull's eye to the total area of the target: $20\%=\pi b^{2}/(\pi R^{2})$.
Therefore $b=R\sqrt{20\%}$. Now, imagine another competition is held, with
all rules the same but the prizes differently assigned to the rings of the
target. Suppose the ten archers got 500\$, 500\$, 100\$, 100\$, 100\$, 50\$,
50\$, 50\$, 50\$, 50\$. Again, we can determine that the 20\% of the highest
rewards correspond to hitting the central spot, and can determine its radius
exactly $b$ as before. Note that in the determination of $b$ we are not
using the actual values of the rewards at all -- the function used can be
any monotonic function of the centrality. The rewards are only used to {\em %
categorize} the data. Once this is done, we can identify the $c$ ``most
central'' archers and determine $b$ according to Eq. (\ref{rel}),
irrespectively of the function used for categorizing. Our example can be
translated into heavy-ion collisions in the following way: archery
competition -- heavy-ion experiment, archer that scored -- event, rewards in
competition I -- number of participants, rewards in competition II --
multiplicity of produced particles, percentile of highest-scoring archers --
centrality, radii of rings on the target -- impact parameters.

The above example shows the essence of our argument, valid for the classical
physics of relativistic heavy-ion collisions. There are, however, two
additional features which need to be considered.
First, a collision at a particular impact parameter $b$ produces values of $%
n $ which are statistically distributed around some mean value $\bar{n}(b)$
with a distribution width $\Delta n(b)$. As we will show, Eq.(\ref{rel}),
formally valid at $\Delta n(b) \ll \bar{n}(b)$, is accurate even for
realistically large $\Delta n(b)$, such as obtained from statistical models
of particle production. Second, there are boundary effects near $b\sim R$
--- at lower values of $b$ the inelastic cross section is the cross section
for colliding {\em black disks}, whereas at the boundary the target
gradually becomes transparent.

We now proceed with a formal derivation. Let $P(n)$ denote the probability
of obtaining value $n$ for the categorizing function (multiplicity of
produced particles, number of participants, number of binary collisions, 
{\em etc.)}. For simplicity of the language we call it the {\em multiplicity}%
, bearing in mind it could be any of these quantities. The centrality $c$ is
defined as the cumulant of $P(n)$, namely 
\begin{equation}
c(N)=\sum_{n=N}^{\infty }P(n).  \label{central}
\end{equation}
Thus $c(N)$ is the probability of obtaining an event with multiplicity
larger or equal to $N$. A particular value of multiplicity $n$ may be
collected from collisions with various impact parameters $b^{\prime }$, thus
we can write 
\begin{equation}
c(N)=\sum_{n=N}^{\infty }\int_{0}^{\infty }\frac{2\pi b^{\prime }db^{\prime }%
}{\sigma _{{\rm inel}}}\rho (b^{\prime })P(n|b^{\prime }),  \label{cen2}
\end{equation}
where $2\pi b^{\prime }db^{\prime }$ is the area of the ring between impact
parameters $b^{\prime }$ and $b^{\prime }+db^{\prime }$, the quantity $\rho
(b^{\prime })$ is the probability of an event (inelastic collision) at
impact parameter $b^{\prime }$, and $P(n|b^{\prime })$ is the conditional
probability of producing multiplicity $n$ provided the impact parameter is $%
b^{\prime }$. The function $\rho (b^{\prime })$ is unity for $b^{\prime }$
below $R$, and drops smoothly to zero at $b^{\prime }$ around $R$,
reflecting the washed-out shape of the nuclear density functions at the
edges. The interpretation of Eq. (\ref{cen2}) is clear: the probabilities
for hitting the ring between $b^{\prime }$ and $b^{\prime }+db^{\prime }$,
the probability for an event to occur at $b^{\prime }$, and the probability
to produce multiplicity $n$ (provided the event occurred at $b^{\prime }$)
are multiplied, as requested by the classical nature of the problem. Since
we have $\sum_{n=1}^{\infty }P(n|b^{\prime })=1$, and, by definition, $%
\int_{0}^{\infty }2\pi b^{\prime }db^{\prime }\rho (b^{\prime })=\sigma _{%
{\rm inel}}$, we verify the proper normalization in Eq. (\ref{cen2}), namely 
$c(1)=1.$ Furthermore, for heavy nuclei we may use the continuity limit, $%
\sum_{n=N}^{\infty }\rightarrow \int_{N}^{\infty }dn=\int_{0}^{\infty
}dn\,\theta (n-N)$.

The function $P(n|b^{\prime })$ is not known, yet, by the statistical nature
of the particle production, and by experience of various models, we expect
that for large values of $n$ it is narrowly peaked around an average value $%
\bar{n}(b^{\prime })$. Thus we begin our study by taking the limit of an
infinitely-narrow distribution, $P(n|b^{\prime })=\delta (n-\bar{n}%
(b^{\prime }))$. In this case 
\begin{eqnarray}
c(N) &=&\int_{0}^{\infty }dn\,\theta (n-N)\int_{0}^{\infty }\frac{2\pi
b^{\prime }db^{\prime }}{\sigma _{{\rm inel}}}\rho (b^{\prime })\delta (n-%
\bar{n}(b^{\prime }))  \nonumber \\
&=&\int_{0}^{\infty }\frac{2\pi b^{\prime }db^{\prime }}{\sigma _{{\rm inel}}%
}\rho (b^{\prime })\theta (\bar{n}(b^{\prime })-N).  \label{cenbis}
\end{eqnarray}
Since $\bar{n}(b^{\prime })$ is a monotonically{\em \ decreasing} function
of $b^{\prime }$, we have $\theta (\bar{n}(b^{\prime })-N)=\theta
(b(N)-b^{\prime })$, where $b(N)$ is the solution of the equation $\bar{n}%
(b)=N$. Therefore 
\begin{eqnarray}
c(N) &=&\int_{0}^{\infty }\frac{2\pi b^{\prime }db^{\prime }}{\sigma _{{\rm %
inel}}}\rho (b^{\prime })\theta (b(N)-b^{\prime })  \nonumber \\
&=&\int_{0}^{b(N)}\frac{2\pi b^{\prime }db^{\prime }}{\sigma _{{\rm inel}}}%
\rho (b^{\prime })=\frac{\sigma _{{\rm inel}}(b(N))}{\sigma _{{\rm inel}}},
\label{cen4}
\end{eqnarray}
where $\sigma _{{\rm inel}}(b(N))$ is the inelastic cross section
accumulated from $b^{\prime }\leq b(N)$. Equation (\ref{cen4}) is a
generalization of formula (\ref{rel}). In Ref. \cite{vogt} it has been quoted in
the context of the Glauber model. We notice that although $c$ and $b$ depend
implicitly on $N$, their relation does not explicitly involve $N$.

We now turn to a quantitative analysis of dispersion effects. Assume 
\begin{equation}
P(n|b^{\prime })=\frac{1}{\Delta n(b^{\prime })\sqrt{2\pi }}\exp \left( -%
\frac{(n-\bar{n}(b^{\prime }))^{2}}{2\Delta n(b^{\prime })^{2}}\right) ,
\label{gauss}
\end{equation}
which is a good approximation for $\Delta n(b^{\prime })<\bar{n}(b^{\prime
}) $. Then 
\begin{equation}
c(N)=\int_{0}^{\infty }\frac{2\pi b^{\prime }db^{\prime }}{\sigma _{{\rm inel%
}}}\rho (b^{\prime })\left\{ \frac{1}{2}\left[ \mathop{\rm erf}\left( \frac{%
\bar{n}(b^{\prime })-N}{\sqrt{2}\Delta n(b^{\prime })}\right) +1\right]
\right\} .  \label{cA}
\end{equation}
For small $\Delta n(b^{\prime })$ the function in curly brackets resembles
the function $\theta (\bar{n}(b^{\prime })-N)$, washed out over the range $%
\Delta n(b^{\prime })$. Thus, we introduce the function 
\begin{equation}
d(x)=\frac{1}{2}\left[ \mathop{\rm erf} \left( \frac{x}{\sqrt{2}\Delta n}%
\right) +1\right] -\theta (x).  \label{F}
\end{equation}
The integral of $d(x)$ with a regular function $f(x)$ can be expanded in
even powers of $\Delta n$ as follows (this is analogous in spirit to the
Sommerfeld expansion of the Fermi-Dirac distribution function at low
temperatures): 
\begin{equation}
\int dx\,f(x)d(x)=-\!\!\! \sum_{j=1,3,5,...}\!\!\!\!a_{j}(\Delta
n)^{j+1}\left. \frac{d^{j}f(x)}{dx^{j}}\right| _{x=0},  \label{some}
\end{equation}
with the coefficients 
\begin{eqnarray}
a_{j} &=&\frac{1}{j!}\int_{-\infty}^{\infty} dx\,x^{j}d(x)=\frac{%
2^{(j+3)/2}\Gamma (\frac{j}{2}+1)}{\sqrt{\pi }(j+1)!},  \label{aj} \\
a_{1} &=&1,\quad a_{3}=\frac{1}{4},\quad a_{5}=\frac{1}{24}, \quad a_{7}=%
\frac{1}{192},\quad ...  \nonumber
\end{eqnarray}
We rewrite the integral in Eq. (\ref{cA}) as $\int 2b^{\prime }db^{\prime
}=\int d\bar{n}\,db^{\prime 2}/d\bar{n}$, and use expansion (\ref{some}) to
obtain 
\begin{eqnarray}
&& c(N) =\frac{\sigma _{{\rm inel}}(b(N))}{\sigma _{{\rm inel}}}
\label{ccorr} \\
&&-(\Delta n(b(N))^{2}\frac{d}{d\bar{n}}\left. \left( \frac{\pi \rho (b(\bar{%
n}))}{\sigma _{{\rm inel}}}\frac{db^{2}(\bar{n})}{d\bar{n}}\right) \right| _{%
\bar{n}=N}-...  \nonumber
\end{eqnarray}
For inner $b$, where $\rho (b(\bar{n}))\simeq 1$, the correction term is
proportional to $d^{2}(b^{2}(\bar{n}))/d\bar{n}^{2}$. In the models
considered below this quantity is proportional to $1/\bar{n}^{2}$, and as a
result $c(N)=\sigma _{{\rm inel}}(b(N))/\sigma _{{\rm inel}}+{\cal O}(\Delta
n^{2}/\bar{n}^{2})$, quantitatively showing that the geometric
identification (\ref{rel}), or (\ref{cen4}), is good for narrow
distributions.

In order to illustrate the above results and to obtain more detailed
numerical estimates for the corrections we consider two models: a model
inspired by the {\em wounded-nucleon} model \cite{bbc}, 
and the optical limit of the 
{\em Glauber model} \cite{wong} 
for the binary collisions. A combination of these models
has been used to explain the observed hadron multiplicities produced in RHIC 
\cite{phenixcentr}. We look at the $Au+Au$ reaction, with the nucleus
density profile $\rho _{A}(r)$ described be the standard Woods-Saxon
function with the radius $r_{0}=(1.12A^{1/3}-0.86A^{-1/3}){\rm fm}$, with $%
A=197$, and the width parameter $a=0.54{\rm fm}$. The nucleus-nucleon
thickness function is given by $T_{A}(s)=\int_{-\infty }^{\infty }dz\,\rho
_{A}(\sqrt{s^{2}+z^{2}})$, and the average number of wounded nucleons is
\begin{eqnarray}
\bar{n}(b) &=&2A\int_{0}^{\infty }sds\int_{0}^{2\pi }d\varphi T_{A}(\sqrt{%
s^{2}+b^{2}+2sb\cos \varphi })\times   \nonumber \\
&&\left( 1-\left( 1-\sigma T_{A}(s)\right) ^{A}\right) ,  \label{nbarw}
\end{eqnarray}
where, following Ref. \cite{phenixcentr}, we take $\sigma =40{\rm mb}$ as
the nucleon-nucleon inelastic cross section. The total nucleus-nucleus cross
section obtained in this model is $\sigma _{{\rm inel}}=7.05{\rm b}$. The
expressions for the dispersion of wounded nucleons produced at a given $b$
is very complicated. Instead of computing multidimensional integrals, we
explore, for our illustrative purpose, two cases: $\Delta n\sim \bar{n}$,
and $\Delta n\sim \sqrt{\bar{n}}$. Led by the sample numerical results for
the distributions given in Fig. 1 of Ref. \cite{bm}, we take {\em i)} $%
\Delta n=\bar{n}/10$, or {\em ii)} $\Delta n=\sqrt{\bar{n}}$. In Fig. 1 we
show the results of computing $c(N)$ according to Eqs. (\ref{cA},\ref{nbarw}%
) with $\rho (b^{\prime })=\theta (\sqrt{\sigma _{{\rm inel}}/\pi }%
-b^{\prime })$, and for the choices {\em i)} and {\em ii)} (dot-dashed and
dashed lines, respectively). These are compared to $\pi b(N)^{2}/\sigma _{%
{\rm inel}}$ (solid line), where $b(N)$ is defined as the solution of the
equation $\bar{n}(b)=N$. The curves overlap within the width of the line,
except for tiny regions at very low $N$, ( $N<2$ ), corresponding to very
peripheral collisions, and at large $N$, corresponding to $b$ around $0$.
The discrepancy at large $N$ follows from the fact that $c(N)$ evaluated
exactly continues to be non-zero till the maximum value of wounded nuclei, $%
N=2A$, whereas $b(N)$ by construction goes to zero at $N=\bar{n}(b=0)\approx
377$. This effect is visible in Fig. 1 only for the choice {\em i)} for the
widths.

We can treat the dependence on $N$ as parametric, and plot $c(b(N)) $ vs. $%
b(N)$. The results is shown in Fig. 2(a). Again, the model curves for $c(b)$
for choices {\em i)} and {\em ii)} overlap with the curve $\pi b^{2}/\sigma
_{{\rm inel}}$ except for very peripheral ($b>14{\rm fm}$) and very central (%
$b<2{\rm fm}$) collisions. This behavior directly reflects the behavior of
Fig. 1. The size of the correction of Eq. (\ref{ccorr}) is, at intermediate $%
b$, of the order of 10$^{-3}$.

\begin{figure}[tbh]
\epsfysize=5.2cm
\centerline{\mbox{\epsfbox{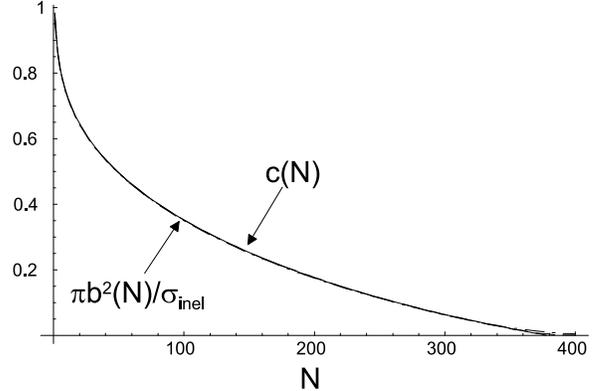}}}
\caption{Centrality in models {\em i)} and {\em ii)} (dot-dashed and dashed
lines), and the function $\pi b(N)^{2}/\sigma_{\rm inel}$
(solid line), plotted as functions of the number of participants, $N$.}
\label{fi1}
\end{figure}

As another illustrative example we consider the Glauber model of
nucleus-nucleus collisions and analyze binary collisions, $n=n_{coll}$. We
use the optical limit of the model, which results in simple expressions. In
this model 
\begin{eqnarray}
&&c\left( N\right) =\sum_{n=N}^{A^{2}}\int_{0}^{\infty }\frac{2\pi b^{\prime
}db^{\prime }}{\sigma _{{\rm inel}}}P_{G}\left( n,b^{\prime }\right) =
\label{cG} \\
&&\sum_{n=N}^{A^{2}}\int_{0}^{\infty }\frac{2\pi b^{\prime }db^{\prime }}{%
\sigma _{{\rm inel}}}\left( 
\begin{array}{c}
A^{2} \\ 
n
\end{array}
\right) \left[ T\left( b^{\prime }\right) \sigma \right] ^{n}\left[
1-T\left( b^{\prime }\right) \sigma \right] ^{A^{2}-n},  \nonumber
\end{eqnarray}
where for $P_{G}\left( n,b^{\prime }\right) \ $we have used the formula for
the probability of the occurrence of $n$ inelastic baryon-baryon collisions
at an impact parameter $b^{\prime }$ \cite{wong} (note that $P_{G}$ plays
the role of the product $\rho (b^{\prime })P(n|b^{\prime })$ from the
previous discussion). Here $T\left( b\right) $ is the nucleus-nucleus
thickness function, 
\begin{eqnarray}
T\left( b\right)  &=&\int_{0}^{\infty }ds\int_{-\infty }^{\infty
}dz_{A}\,\int_{-\infty }^{\infty }dz_{B}\int_{0}^{2\pi }d\varphi \times 
\label{T} \\
&&\,\rho (\sqrt{s^{2}+z_{A}^{2}})\rho (\sqrt{s^{2}+b^{2}+2sb\cos \varphi
+z_{B}^{2}}).  \nonumber
\end{eqnarray}
The sum in Eq. (\ref{cG}) can be carried out exactly, yielding, with the
notation $x=T\left( b^{\prime }\right) \sigma $, the expression 
\begin{eqnarray}
c\left( N\right)  &=&\int_{0}^{\infty }\frac{2\pi b^{\prime }db^{\prime }}{%
\sigma _{{\rm inel}}}\left( 
\begin{array}{c}
A^{2} \\ 
n
\end{array}
\right) \times   \label{hyper} \\
&&(1-x)^{A^{2}}x^{N}\,_{2}F_{1}(1,N-A^{2};N+1;\frac{x}{x-1}).  \nonumber
\end{eqnarray}
We perform the integration in Eq. (\ref{hyper}) numerically. On the other
hand, the average number of collisions at a fixed value of the impact
parameter $b$ is $\bar{n}\left( b^{2}\right) =A^{2}\,T\left( b\right) \sigma
$. Repeating the steps of the previous example results in the identification
$c(N)=c(\bar{n}\left( b^{2}\right) )=c(A^{2}\,T\left( b\right) \sigma )$.
This function is compared to $\pi b^{2}/\sigma _{{\rm inel}}$ in Fig. 2(b).
The agreement is excellent and the two curves are indistinguishable except
for very peripheral collisions ($b>13.5$fm).

\begin{figure}[tb]
\epsfysize=8.2cm
\centerline{\mbox{\epsfbox{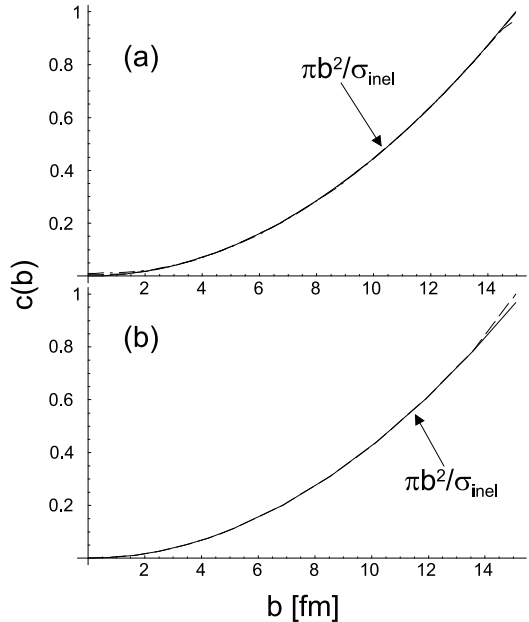}}~~~~}
\caption{(a) Centrality as a function of the impact parameter for the models
{\em i)} (dot-dashed line) and {\em ii)} (dashed line). (b) The same for the
Glauber model for binary collisions (dashed line). Solid line shows 
$\pi b^{2}/\sigma_{{\rm inel}}$.}
\label{f2}
\end{figure}

With the Gaussian parametrization of the thickness function the model can be
treated analytically a bit further. We use $T\left( b^{\prime \,}\right)
=1/(2\pi \beta ^{2})\exp \left( -b^{\prime \,2}/(2\beta ^{2})\right) $, with 
$\beta =4.6$fm, which leads to a quite good approximation of the exact
thickness function. Then 
\begin{eqnarray}
c\left( N\right)  &=&\frac{2\pi \beta ^{2}}{\sigma _{{\rm inel}}}%
\sum_{n=N}^{A^{2}}\left( 
\begin{array}{c}
A^{2} \\ 
n
\end{array}
\right) \int_{0}^{\sigma /(2\pi \beta ^{2})}\!\!\!\!\!\!\!dy\,y^{n-1}\left(
1-y\right) ^{A^{2}-n}  \nonumber \\
&=&\frac{2\pi \beta ^{2}}{\sigma _{{\rm inel}}}\sum_{n=N}^{AB}\frac{1}{n}%
I_{\sigma /(2\pi \beta ^{2})}\left( n,1+A^{2}-n\right) ,  \label{cGcal}
\end{eqnarray}
where$\,I_{z}\left( a,b\right) =B_{z}\left( a,b\right) /B_{1}\left(
a,b\right) \,$, and $B_{z}\left( a,b\right) $ is the incomplete beta
function. For large $A^{2}$ and small $\sigma /(2\pi \beta ^{2})$ the
function$\,I_{\sigma /(2\pi \beta ^{2})}\left( n,1+A^{2}-n\right) $ is well
approximated by the step function $\theta \left( A^{2}\sigma /(2\pi \beta
^{2})-n\right) $. Replacing the sum by the integral in (\ref{cGcal}) we find
the leading expression 
\begin{equation}
c\left( N\right) =-\frac{2\pi \beta ^{2}}{\sigma _{{\rm inel}}}\ln \left( 
\frac{2\pi \beta ^{2}}{AB\sigma }N\right) .  \label{cGofN}
\end{equation}
On the other hand 
$b^{2}\left(
N\right) =-2\beta ^{2}\ln \left( 2\pi \beta ^{2}N/(AB\sigma )\right)$,
which immediately results in Eq. (\ref{rel}). Since $\Delta
n^{2}=A^{2}T(b)\sigma (1-T(b)\sigma )\simeq \bar{n}$, the correction of Eq. (%
\ref{ccorr}) becomes $-2\pi \beta ^{2}/\sigma _{{\rm inel}}(\Delta
n/N)^{2}\simeq -2\pi \beta ^{2}/\sigma _{{\rm inel}}(1/N)\simeq -0.2/N$,
hence is very small at large $N$.

As already mentioned, there are attempts \cite{phenixcentr} to explain the
multiplicity of produced particles through a combination of the {\em wounded}
nucleon model \cite{bbc}, 
associated with soft processes, and production proportional
to the number of binary nucleon-nucleon collisions, associated with hard
physics. The folding of the distributions of wounded nucleons, $n_{w}$,
or number of collisions, $n_{{\rm coll}}$,
with the distribution of particles produced in an elementary event (by the
wounded nucleon or in a single binary collision), may result in a
broadening effect in the observed distribution of the multiplicity of 
the produced particles, $n$. However,
we expect this broadening to be negligible in the ratio $\Delta n/%
\bar{n}$, wich is the quantity controlling the accuracy of Eq. (\ref{rel}). 
In particular, for the wounded nucleon model \cite{bm} one has $%
(\Delta n/\bar{n})^{2}=2(\Delta _{H})^{2}/(\bar{n}_{w}\bar{n}%
_{H}^{2})+(\Delta n_{w}/\bar{n}_{w})^{2}$, where the subscript ${}_{H}$
refers to the nucleon-nucleon collision. Assuming $(\Delta _{H})^{2}\sim 
\bar{n}_{H}$, we find that the contribution from 
the first term is smaller than from the second term
already for moderately large $\bar{n}_{w}$, and $\Delta n/\bar{n}\simeq
\Delta n_{w}/\bar{n}_{w}$. This indicates that Eq. (\ref{rel}) remains very
accurate when multiplicities of produced particles are used as the
centrality criterion.

We wish to thank Andrzej Bia\l{}as, Andrzej Bu\-dza\-nowski, Wies\l{}aw
Czy\.z, Roman Ho\l{}y\'nski, Pasi Huovinen, and Kacper Zalewski for useful
discussions.

\end{document}